\documentclass{article}

\usepackage{arxiv}

\usepackage[utf8]{inputenc}
\usepackage[T1]{fontenc}
\usepackage{hyperref}
\usepackage{url}
\usepackage{booktabs}
\usepackage{amsfonts}
\usepackage{nicefrac}
\usepackage{microtype}
\usepackage{graphicx}
\usepackage{amsmath,amssymb}
\usepackage{bm}
\usepackage{mathrsfs}
\usepackage{amsthm}
\usepackage{multirow}
\usepackage{float}
\usepackage{xcolor}

\newtheorem{theorem}{Theorem}
\newtheorem{corollary}{Corollary}
\newtheorem{proposition}{Proposition}

\newcommand{\ZZ}{\mathbb{Z}_2}
\newcommand{\avg}[1]{\left\langle #1 \right\rangle}
\newcommand{\dd}{\mathrm{d}}
\newcommand{\ee}{\mathrm{e}}
\newcommand{\ii}{\mathrm{i}}
\newcommand{\ndot}{\dot{n}}
\newcommand{\Phisec}{\hat{\Pi}_{\Phi}}
\newcommand{\Gm}[1]{\hat{G}_{m_{#1}}}
\newcommand{\nbar}[1]{\bar{n}_{m_{#1}}}
\newcommand{\Psip}{|\Psi^{+}_{\ell_1\ell_2}\rangle}
\newcommand{\Phip}{|\Phi^{+}_{\ell_1\ell_2}\rangle}
\newcommand{\Ucal}{\mathcal{U}}
\newcommand{\Kcal}{\mathcal{K}}
\newcommand{\Lcal}{\mathcal{L}}
\newcommand{\tr}{\mathrm{tr}\,}
\newcommand{\Pnom}{\hat{\Pi}_{\rm nom}}

\title{Recovering What Breaks in a Trapped-Ion Quantum Simulation of a Lattice Gauge Theory}

\author{
  Hassan Ugail \\
  Centre for Visual Computing and Intelligent Systems \\
  University of Bradford \\
  United Kingdom \\
   \\
}

\begin{document}
\maketitle

\begin{abstract}
A recent trapped-ion experiment realised a lattice gauge theory with the gauge field in the internal states of the ions and bosonic matter in their vibrations. We showed earlier that its noise damage can be predicted. Here, we show that most of it can be undone. The conservation law of the theory is a parity of each vibrational mode, which heating flips, so a heating event leaves a mark that can be read without disturbing the physics. We keep only the runs in which this mark never changed. With equal heating rates and continuous monitoring, the kept runs reproduce the heating-free dynamics exactly, with a closed-form yield. Reading the mark at finite intervals introduces a small bias linear in the interval, which we derive and bound. At the reported rates, a single final check removes most of the heating and preparation damage while accepting eight runs out of ten.
\end{abstract}


\section{Introduction}
\label{sec:intro}

A lattice gauge theory ties the motion of matter to a field that lives on the links between lattice sites.
The local conservation law that expresses this tie is Gauss's law, and a quantum simulation of a gauge theory
is judged by how well it respects it~\cite{wiese,dalmonte,banuls}. Over the past decade, the simplest discrete
example, the $\ZZ$ lattice gauge theory coupled to dynamical matter, has become a laboratory object. It has
been realised with trapped ions~\cite{martinez}, with ultracold atoms in Floquet-engineered double
wells~\cite{schweizer}, and with superconducting qubits in which both matter and gauge field are
spins~\cite{mildenberger}. Most recently, it has been realised on a trapped-ion processor in which the gage
field is encoded in the internal state of each ion and the matter in the quantised vibrations of the
crystal~\cite{bazavan,saner}. In that experiment, two ions close a two-link loop. The link qubits are prepared
in Bell states that fix a $\ZZ$ flux through the loop, and a phonon either tunnels freely or is frozen in
place by Aharonov--Bohm interference~\cite{aharonov}, depending on the flux.

The hybrid qubit--oscillator encoding gives the matter a feature that spin and hard-core encodings lack. A site
can hold any number of excitations, and the Gauss generator at a site is the product of the phonon parity of that site and the link qubits that meet it. The local conservation law is therefore a parity, and a parity can
be measured without learning anything else about the state. Three noise processes dominate the device. The
vibrational modes heat, their motional coherence decays, and the qubits suffer errors. In a companion paper, we
showed that the damage each process does to each observable class of the experiment can be predicted before
the machine is run, as certified bounds and first-order laws derived from the reported noise
rates~\cite{ugail2026noise}. The central fact of that paper is that heating violates Gauss's law by flipping a
parity at a rate that grows with the number of quanta already in the mode.

The present paper turns that fact around. A heating event flips a parity, and the parity can be read without
disturbing the gauge-invariant dynamics, so a heating event leaves a mark. One can read the mark at intervals
and keep only the runs in which it never changed. This is symmetry verification by post-selection, and it has
a substantial literature. It was proposed as a low-cost error mitigation strategy for variational
algorithms~\cite{bonetmonroig,mcardle} and has since been applied to gauge theories in several forms. Sparse
error detection with a final post-selection on the gauge charges has been studied for Hamiltonian
simulation~\cite{froland}. Dynamical post-selection on mid-circuit Gauss-law checks has been worked out for
non-Abelian theories on qudits~\cite{ballini}. Full stabiliser codes built from Gauss's law have been
constructed for hard-core $\ZZ$ matter~\cite{modi}. The older idea of pinning the dynamics to the
gauge-invariant subspace by an engineered measurement or dissipation is due to~\cite{stannigel}, and the
reliability of the gauge-invariant dynamics under coherent and incoherent errors has been analysed in
general~\cite{halimeh2020reliability,halimeh2020fate,halimeh2021protection}. All of this work treats matter
that is a spin or a hard-core boson. For such matter under Pauli noise, the jump operators are unitary. The
evolution conditioned on no error is then the ideal evolution by construction, and post-selection removes
every detected error exactly. For bosonic matter, the jump operators are not unitary. The evolution conditioned
on no error carries a backaction that in general distorts the surviving state, and post-selection on a symmetry
is then only approximate, however well the errors are detected.

The first result of this paper is that for bosonic matter under heating, the backaction is nevertheless
trivial, so the post-selected estimator is exact. The mechanism itself is not new. A dissipative channel whose
total jump rate depends only on a conserved quantity produces a common decay factor on each sector of that
quantity and leaves the normalised state inside the sector untouched. In the bosonic-code literature, this is
why two-mode codes with a fixed total excitation number are unaffected by the no-jump evolution of uniform
loss~\cite{chuang}, and the construction has recently been formalised as uniformly decaying subspaces for
error-mitigated computation~\cite{suri}. What this paper adds is the recognition that the gauge-invariant
dynamics of the ion platform supplies such a conserved quantity, the total phonon number, and that the Gauss
generators supply a syndrome that reads the heating history without disturbing it. In essence, they give an exact
post-selected estimator with a closed-form yield on any geometry and at any electric field. The result holds
for any heating model in which the sum of the gain and loss rates is the same on every mode.

The rest of the paper works out what happens when the ideal conditions are relaxed, because the experiment
will relax them, and this is where the main analytic content lies. The syndrome cannot be read continuously.
A finite reading interval admits runs in which two heating events on the same mode cancelled between readings.
Such return events are excluded by hypothesis in the uniformly decaying subspace construction~\cite{suri}. The
heating channel produces them at any occupation because an empty mode can gain a quantum and lose it again,
whereas pure loss needs at least two quanta on a mode. We show that the resulting bias grows linearly in the
interval. We derive its coefficient and the accompanying excess yield from the heating-free dynamics alone, and
we give a certificate for the bias of any bounded observable. For a state of definite phonon number the
certificate needs only the measured fraction of accepted runs and the closed-form yield. The heating rates of
the modes are not equal in practice, and unequal rates tilt the conditional state by an amount with a closed
form at zero electric field. The matter is prepared on a thermal background rather than in a Fock state. A
syndrome measurement handles this too, because projecting on the nominal syndrome fixes the parity of every
mode and so removes most of the thermal background along with the heating. Finally, we state what a
destructive Gauss-law measurement at the end of each run can test, with the readout it requires, and what
would need a non-demolition measurement of the generators during the run. The methodology follows the
certification programme we developed for spectral diagnostics in numerical relativity~\cite{ugail2026weyl}
and for noisy quantum neural networks~\cite{ugail2026access,ugail2026coherence}. Every predictive law is
accompanied by a certified bound, and every bound is tested on an exact computation.

\section{Model, noise and the Gauss-law syndrome}
\label{sec:model}

\begin{figure}
\centering
\includegraphics[width=0.98\textwidth]{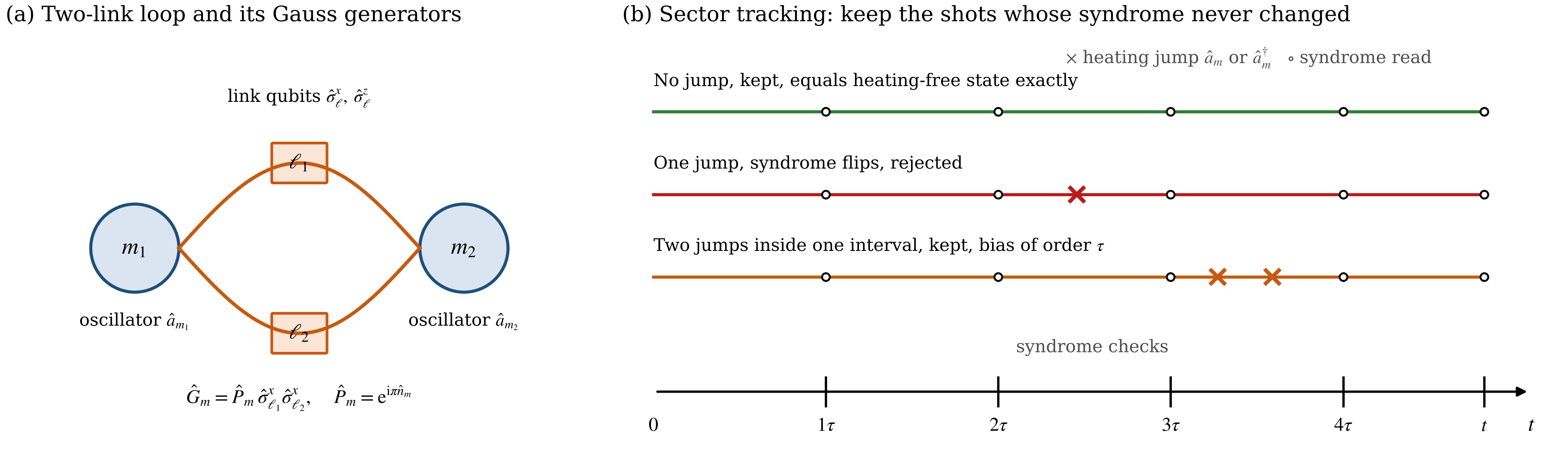}
\caption{Sector tracking on the two-link loop. (a) Two sites $m_1$ and $m_2$ carry oscillators and two links
$\ell_1$ and $\ell_2$ carry qubits. The Gauss generator at each site is the phonon parity of that site
multiplied by the two link qubits in the electric basis. (b) Three trajectories of the quantum-jump
unravelling. The syndrome, which is the pair of generator eigenvalues, is read at intervals $\tau$. A run
with no heating jump is kept and equals the heating-free dynamics exactly. A run with one jump since the last
reading has a flipped syndrome and is rejected. A run with two jumps on the same mode inside one interval has
an unchanged syndrome and is kept. Runs of the last kind are the source of a bias that is linear in $\tau$.}
\label{fig:schematic}
\end{figure}

We use the model and notation of~\cite{ugail2026noise}, which follow~\cite{bazavan,saner}. A graph has sites
$m$ carrying oscillators with ladder operators $\hat a_m$, and links $\ell=(i,j)$ carrying qubits with Pauli
operators $\hat\sigma^{x}_\ell$, $\hat\sigma^{y}_\ell$ and $\hat\sigma^{z}_\ell$. The Hamiltonian of the $\ZZ$
lattice gauge theory on this graph is
\begin{equation}
\hat H=\sum_{\ell=(i,j)}\Big[J\big(\hat a_j^{\dagger}\hat\sigma^{z}_\ell\hat a_i+\mathrm{h.c.}\big)
+h\,\hat\sigma^{x}_\ell\Big],
\label{eq:H}
\end{equation}
where $J$ is the gauge-invariant tunnelling and $h$ is the electric field. A phonon hops between sites only
through the link qubit, which it flips in the electric basis. The generator of the local symmetry at site $m$
is
\begin{equation}
\hat G_m=\hat P_m\prod_{\ell\ni m}\hat\sigma^{x}_\ell,\qquad \hat P_m=\ee^{\ii\pi\hat n_m},
\label{eq:G}
\end{equation}
with $\hat n_m=\hat a_m^{\dagger}\hat a_m$, and $[\hat H,\hat G_m]=0$ at every site. The eigenvalue
$g_m=\pm1$ labels the background charge at $m$. Gauss's law is the statement that the dynamics stays in the
sector fixed by the initial state. The flux through a closed loop $C$ is measured by $\hat
W_C=\prod_{\ell\in C}\hat\sigma^{z}_\ell$. On the two-link loop of the experiment, preparing the link qubits in
$\Phip$ fixes the flux to zero, and preparing them in $\Psip$ fixes it to $\pi$. We write
$\Phisec=\tfrac12(1+\hat\sigma^{z}_{\ell_1}\hat\sigma^{z}_{\ell_2})$ for the projector onto the zero-flux
sector. The Hamiltonian conserves the total phonon number $\hat N=\sum_m\hat n_m$, since each hopping term
moves one phonon between sites and the electric term acts on the qubits alone.
Figure~\ref{fig:schematic}(a) shows the loop and its generators.

The device noise is modelled by a Lindblad master equation $\dot\rho=-\ii[\hat H,\rho]+\sum_k\mathcal D[\hat
L_k]\rho$ with $\mathcal D[\hat L]\rho=\hat L\rho\hat L^{\dagger}-\tfrac12\{\hat L^{\dagger}\hat
L,\rho\}$~\cite{lindblad,gks,breuer}. Motional heating of mode $m$ at $\ndot_m$ quanta per second is modelled
as an infinite-temperature bath with jump operators $\sqrt{\ndot_m}\,\hat a_m$ and $\sqrt{\ndot_m}\,\hat
a_m^{\dagger}$. This is the standard model for electric-field noise on a trap
electrode~\cite{turchette,brownnutt}. Motional dephasing with coherence time $t_{c,m}$ is modelled by the jump
operator $\sqrt{2/t_{c,m}}\,\hat n_m$. Qubit errors on link $\ell$ are modelled as bit flips
$\sqrt{\gamma_x}\,\hat\sigma^{x}_\ell$ and as dephasing $\sqrt{\gamma_z}\,\hat\sigma^{z}_\ell$. For the two
radial modes used on the loop the experiment quotes heating rates of $5$ and $17$ quanta per second, coherence
times of $2.3$ and $3.3$ milliseconds, an initial thermal occupation $\bar n_0=0.1$ on every mode, and a loop
tunnelling rate $J/\pi=0.35$ kilohertz~\cite{saner}. It does not quote qubit error rates. We use the reported
values in Section~\ref{sec:readout}. Elsewhere, we raise the heating rates to $150$ quanta per second, so that
the heating damage within the experimental window of $2.32$ milliseconds is large enough for the recovery to be
visible. Nothing in the analysis depends on the rates being small.

The role of each channel with respect to Gauss's law was established in~\cite{ugail2026noise}, and the method
rests on it, so we restate it. Motional dephasing is a function of the phonon number and commutes with every
generator. Qubit bit flips act with $\hat\sigma^{x}$ and also commute with every generator. Heating adds or
removes one phonon and flips the parity $\hat P_m$ of the mode it acts on, which flips $\hat G_m$ and no other
generator. Qubit dephasing acts with $\hat\sigma^{z}_\ell$, which anticommutes with $\hat\sigma^{x}_\ell$, so
it flips the generators at both ends of the link. We call the vector of generator eigenvalues $(g_m)_m$ the
syndrome. A heating event on mode $m$ changes the syndrome at $m$ alone. A qubit dephasing event on link
$(i,j)$ changes it at $i$ and $j$ together. The syndrome is therefore a record of the heating and
qubit-dephasing history of a run. When the initial state is an eigenstate of every generator, it carries no
information about the gauge-invariant dynamics, because the Hamiltonian commutes with all of them. Reading the
syndrome collapses nothing that the experiment wants to measure.

Sector tracking is the protocol of reading the syndrome at times $\tau,2\tau,\dots$ up to the final time $t$
and keeping only the runs in which it equals the nominal syndrome of the prepared state at every reading.
Figure~\ref{fig:schematic}(b) illustrates the three kinds of run that matter. The observable of interest is
estimated on the kept runs, and the fraction of kept runs is the yield. Throughout the paper, the post-selected
estimator is judged against the gauge-invariant dynamics under the channels the syndrome does not see, which
are motional dephasing and qubit bit flips. We call this the heating-free reference. Qubit dephasing is seen
by the syndrome and would be removed along with the heating. We set $\gamma_z=0$ in every calculation, in line
with the experiment, so that the reference is unambiguous. The effect of the remaining channels on each
observable is given by the certificates of~\cite{ugail2026noise}.

All numerical results are exact Lindblad evolutions with no Monte Carlo sampling. For a state of definite
initial syndrome the post-selected estimator is computed by attaching a parity-flag register to each mode,
which turns it into a ratio of two expectation values on an enlarged space. For the thermal preparation of
Section~\ref{sec:unequal} it is computed by projecting onto the nominal syndrome at every reading, which is
what the physical measurement does. Both constructions, the convergence checks and the truncations, are
described in Appendix~\ref{app:numerics}. We work in units where $J=1$ and quote times in milliseconds using
$J/\pi=0.35$ kilohertz.

\section{Exact recovery under heating}
\label{sec:exact}

The quantum-jump unravelling of a Lindblad equation decomposes the state at time $t$ into contributions from
trajectories with a definite history of jumps~\cite{dalibard,plenio}. The contribution from trajectories with
no heating jump in $[0,t]$ is $\Ucal_{t,0}\rho_0$, where $\Ucal$ is generated by the no-jump Lindbladian
\begin{equation}
\Lcal_{\rm nj}\rho=\Lcal_0\rho-\tfrac12\{\hat K,\rho\},\qquad
\hat K=\sum_{m,\pm}\hat L_{m,\pm}^{\dagger}\hat L_{m,\pm}.
\label{eq:nojump}
\end{equation}
Here $\Lcal_0$ is the Lindbladian of the heating-free reference, and $\hat L_{m,-}$ and $\hat L_{m,+}$ are
the loss and gain jump operators of mode $m$. The expectation of $\hat K$ is the total rate of heating events.
The trace of $\Ucal_{t,0}\rho_0$ is the probability that no heating event occurred in $[0,t]$, and the
normalised state $\Ucal_{t,0}\rho_0/\tr\Ucal_{t,0}\rho_0$ is the state conditioned on that history. The first
result of the paper is that under one condition, this conditional state carries no trace of the heating at
all. We state it for a general bath with loss rate $\kappa_{m,-}$ and gain rate $\kappa_{m,+}$ on mode $m$,
so that $\hat L_{m,-}=\sqrt{\kappa_{m,-}}\,\hat a_m$ and $\hat L_{m,+}=\sqrt{\kappa_{m,+}}\,\hat
a_m^{\dagger}$, because the condition is then visible. The heating model of Section~\ref{sec:model} is the
case $\kappa_{m,-}=\kappa_{m,+}=\ndot_m$.

\begin{theorem}[Exact recovery for a uniform bath]
\label{thm:exact}
Let the sum of the loss and gain rates be the same on every mode, $\kappa_{m,-}+\kappa_{m,+}=s$, and let
$\Lcal_0$ commute with the total phonon number in the sense that $\Lcal_0(\hat N\rho)=\hat N\Lcal_0(\rho)$
and $\Lcal_0(\rho\hat N)=\Lcal_0(\rho)\hat N$. Then on any graph, at any electric field and for any initial
state,
\begin{equation}
\Ucal_{t,0}\rho_0=\ee^{-Ct}\,\ee^{-st\hat N/2}\big(\ee^{t\Lcal_0}\rho_0\big)\ee^{-st\hat N/2},
\qquad C=\sum_m\kappa_{m,+}.
\label{eq:exact}
\end{equation}
If $\rho_0$ lies in the sector of $N$ phonons, the conditional state is $\ee^{t\Lcal_0}\rho_0$ exactly, and
the probability of no heating event is $\ee^{-(sN+C)t}$. For the heating model with equal rates $\ndot$ on
$M$ modes this probability is $\ee^{-\ndot(2N+M)t}$.
\end{theorem}

The proof can be stated concisely. The rate operator is $\hat K=\sum_m(\kappa_{m,-}+\kappa_{m,+})\hat n_m+\sum_m\kappa_{m,+}$,
which under the hypothesis equals $s\hat N+C$. The anticommutator term in Eq.~\eqref{eq:nojump} is then
$-C\rho-\tfrac{s}{2}(\hat N\rho+\rho\hat N)$. The map $\rho\mapsto\hat N\rho+\rho\hat N$ commutes with
$\Lcal_0$ by hypothesis, so the two parts of $\Lcal_{\rm nj}$ can be exponentiated separately, which gives
Eq.~\eqref{eq:exact}. On a sector of fixed $N$ the operator $\hat N$ is the number $N$. The exponential
factors become $\ee^{-(sN+C)t}$, and the normalised state is $\ee^{t\Lcal_0}\rho_0$. The hypothesis on
$\Lcal_0$ holds for the model of Section~\ref{sec:model}. The Hamiltonian commutes with $\hat N$, the motional
dephasing jumps $\hat n_m$ commute with $\hat N$, and the qubit jumps act on a different tensor factor.

Three remarks clarify what the theorem does and does not claim. First, the result is not perturbative. The
conditional state equals the heating-free state at every order in $\ndot t$, and the only effect of the
heating on the kept runs is a number. What makes this work is that the total jump rate $\hat K$ depends on
the conserved quantity $\hat N$ alone, which requires the occupation-dependent part of the rate,
$\kappa_{m,-}+\kappa_{m,+}$, to be the same on every mode. Equal gain and loss within a mode is not required.
Uniform pure loss, uniform pure gain, and a uniform bath at any finite temperature all satisfy the condition.
The first of these is why two-mode bosonic codes with a fixed total excitation number are untouched by the
no-jump evolution of amplitude damping~\cite{chuang}. The same observation underlies the uniformly decaying
subspaces of~\cite{suri}, of which a fixed phonon-number sector of the gauge theory under uniform heating is an
instance. What breaks the condition is a mode dependence of the rates, and Section~\ref{sec:unequal} quantifies
the consequence. Single-mode binomial and cat codes~\cite{michael} face a different problem. Their code words
are superpositions of different photon numbers, on which $\ee^{-st\hat N/2}$ is not a constant.

Second, the theorem concerns the conditional state and not yet a protocol. A protocol reads the syndrome at
intervals and keeps the runs whose syndrome never changed. Those runs include trajectories in which two
heating events on the same mode cancelled between readings. A return event of this kind needs two events on
the same mode inside one interval. For pure loss the mode must hold at least two quanta, since $[\hat
G_m,\hat a_m^2]=0$, so return events are absent for the one-phonon states considered here and present at
higher occupation. For heating, they occur at any occupation, because an empty mode can gain a quantum and
lose it again. The conditional state is therefore only the limit of the protocol as the interval goes to
zero. The construction of~\cite{suri} excludes return events by hypothesis. Section~\ref{sec:finite}
quantifies the approach to the limit, and that part of the analysis is what the heating channel makes
necessary. Third, the thermal preparation of the experiment does not have a definite phonon number, and
Section~\ref{sec:unequal} quantifies that consequence too.

A corollary concerns qubit dephasing. We include it for completeness, although the numerics set $\gamma_z=0$.
A dephasing jump $\sqrt{\gamma_z}\,\hat\sigma^{z}_\ell$ flips the syndrome at both ends of its link, so the
protocol rejects it. Since $\hat\sigma^{z\dagger}\hat\sigma^{z}=1$, its no-jump backaction is the constant
$\ee^{-\gamma_z t/2}$ on every state. Sector tracking therefore removes qubit dephasing exactly as well in the
limit of continuous monitoring, at a further cost of $\ee^{-\gamma_z t}$ per link in the yield. At a finite
interval, two dephasing events on the same link inside one interval restore the syndrome and contribute a bias
of the kind analysed in Section~\ref{sec:finite}. For a device with $\gamma_z\neq0$ the reference is the
dynamics with that channel removed. Bit flips commute with the generators, are invisible to the syndrome, and
are not removed. Their effect on the flux and on the frozen sector is given by the exact exponential laws
of~\cite{ugail2026noise}.

\begin{corollary}[Closed-form yield]
\label{cor:yield}
For an initial state that is a mixture over phonon-number sectors with weights $p_N$, the probability of no
heating event in $[0,t]$ under the heating model with equal rates is $Y_0(t)=\sum_N
p_N\,\ee^{-\ndot(2N+M)t}$. For a state of definite $N$, it is $\ee^{-\ndot(2N+M)t}$.
\end{corollary}

The yield depends on the number of modes even when they are empty because an empty mode can gain a quantum.
For one phonon on the two modes of the loop, it is $\ee^{-4\ndot t}$, and for one phonon on three modes, it is
$\ee^{-5\ndot t}$. A finite-temperature bath would reduce this cost. The infinite-temperature bath of the trap
electrode does not.

\begin{figure}
\centering
\includegraphics[width=0.98\textwidth]{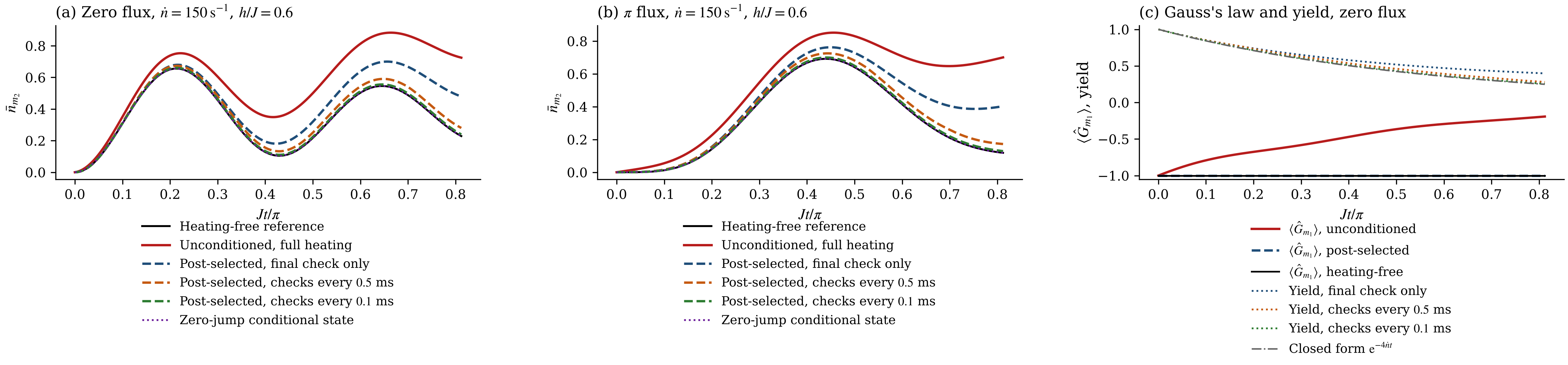}
\caption{Sector tracking on the two-link loop with one phonon, equal heating rates of $150$ quanta per second,
the experimental coherence times and $h/J=0.6$. (a) Population of the second site in the zero-flux sector.
The unconditioned evolution departs from the heating-free reference by up to $0.50$ within the window. The
post-selected estimator approaches the reference as the syndrome interval shrinks, and the zero-jump
conditional state lies on top of the reference. (b) The same in the $\pi$-flux sector, where the heating
damage is $0.58$, and the recovery is the same. (c) The Gauss generator at the first site and the yield in the
zero-flux sector. The post-selected generator is restored to its initial value to machine precision at every
syndrome interval, and the yield approaches the closed form $\ee^{-4\ndot t}$ of
Corollary~\ref{cor:yield} as the interval shrinks.}
\label{fig:loop}
\end{figure}

Figure~\ref{fig:loop} shows the theorem at work on the loop. The initial state is one phonon on the first site
with the link qubits in $\Phip$ or $\Psip$. The heating rates are $150$ quanta per second on both modes, the
coherence times are the experimental ones, and the electric field is $h/J=0.6$. In the zero-flux sector, the
unconditioned population of the second site departs from the heating-free reference by up to $0.50$ within the
window, and in the $\pi$-flux sector by up to $0.58$. The zero-jump conditional state, computed directly from
Eq.~\eqref{eq:nojump}, agrees with the reference to $2.4\times10^{-4}$ in the zero-flux sector and to
$1.5\times10^{-4}$ in the $\pi$-flux sector at the default integrator tolerances on the flagged space. The
generators are conserved to $10^{-15}$. The residual on the populations is numerical. Tightening the
tolerances by two and by four orders of magnitude reduces it to $6\times10^{-7}$ and to $1\times10^{-8}$ at
both Fock truncations tested, and the yield then agrees with the closed form $\ee^{-4\ndot t}=0.2486$ to
$10^{-12}$ and to $10^{-15}$. The convergence is reported in Appendix~\ref{app:numerics}.

\begin{table}
\centering
\caption{Theorem~\ref{thm:exact} on three geometries with one phonon, equal heating rates of $150$ quanta per
second, the experimental coherence times and $h/J=0.3$. The damage is the largest departure of the
unconditioned evolution from the heating-free reference within the window, over all site populations or all
generators. The error is the same quantity for the zero-jump conditional state at the default integrator
tolerances. The yield is the probability of no heating event at the end of the window, and its closed form is
$\ee^{-\ndot(2N+M)t}$ with $N=1$ and $M=3$.}
\label{tab:geometries}
\small\setlength{\tabcolsep}{5pt}
\begin{tabular}{lcccccc}
\toprule
 & \multicolumn{2}{c}{Unconditioned damage} & \multicolumn{2}{c}{Zero-jump error} & \multicolumn{2}{c}{Yield} \\
\cmidrule(lr){2-3}\cmidrule(lr){4-5}\cmidrule(lr){6-7}
Geometry & Populations & Generators & Populations & Generators & Computed & Closed form \\
\midrule
Chain, fluxes $(0,\pi)$ & $0.37$ & $0.75$ & $1.8\times10^{-5}$ & $1.1\times10^{-15}$ & $0.17552$ & $0.17552$ \\
Chain, fluxes $(0,0)$ & $0.41$ & $0.75$ & $8.3\times10^{-6}$ & $8.9\times10^{-16}$ & $0.17552$ & $0.17552$ \\
Triangle, flux $\pi$ & $0.33$ & $0.012$ & $1.4\times10^{-6}$ & $8.5\times10^{-17}$ & $0.17552$ & $0.17552$ \\
\bottomrule
\end{tabular}
\end{table}

The theorem is free of geometry. Table~\ref{tab:geometries} confirms it on a two-loop chain of three sites and
four links in the flux sectors $(0,\pi)$ and $(0,0)$, and on a $\pi$-flux triangle of three sites and three
links. In every case, the zero-jump conditional state reproduces the heating-free populations to a few parts
in a million at the default tolerances, reproduces the generators to machine precision, and has the
closed-form yield to nine digits. These runs use a Fock space of dimension two per mode for the conditional
state and four for the unconditioned damage. Dimension two is exact for the conditional state because the
no-jump dynamics conserves the total phonon number and the initial state has one phonon.

\section{Finite syndrome intervals}
\label{sec:finite}

The syndrome is read at times $k\tau$ for $k=1,\dots,t/\tau$, and a run is kept when every reading equals the
nominal syndrome. For a state of definite initial syndrome, a heating event on mode $m$ flips $g_m$, so a run
is kept exactly when the number of heating events on every mode inside every interval is even. The kept
ensemble therefore contains the zero-jump trajectories, with weight $Y_0(t)$, together with trajectories in
which two or more events on one mode cancelled inside an interval. When the rates are uniform and the initial
state has a definite phonon number, the zero-jump state is the heating-free state by
Theorem~\ref{thm:exact}. Under those two hypotheses, and writing $Y(t)$ for the total weight of the kept
ensemble, which is the yield the experiment measures, the kept state is
\begin{equation}
\rho_{\rm kept}=\frac{Y_0}{Y}\,\rho_{\rm ref}+\Big(1-\frac{Y_0}{Y}\Big)\rho_{\rm rest},
\label{eq:split}
\end{equation}
where $\rho_{\rm rest}$ is the normalised state of the kept trajectories that had at least one cancelled pair.
Everything that follows concerns the second term.

\begin{theorem}[Yield-excess certificate]
\label{thm:cert}
Let the heating rates be uniform, and the initial state have a definite phonon number and a definite syndrome,
so that the zero-jump state equals the reference, and let $\hat O$ be a bounded observable with range $R$,
the difference between its largest and smallest eigenvalue. Then, for every syndrome interval $\tau$ and every
heating rate,
\begin{equation}
\big|\tr\hat O\rho_{\rm kept}-\tr\hat O\rho_{\rm ref}\big|\le R\Big(1-\frac{Y_0(t)}{Y(t)}\Big),
\label{eq:cert}
\end{equation}
where $Y_0$ is the closed form of Corollary~\ref{cor:yield} and $Y$ is the measured yield.
\end{theorem}

The proof follows at once from Eq.~\eqref{eq:split}, since $\tr\hat O\rho_{\rm kept}-\tr\hat O\rho_{\rm
ref}=(1-Y_0/Y)(\tr\hat O\rho_{\rm rest}-\tr\hat O\rho_{\rm ref})$ and the bracket is bounded by $R$. The
certificate is exact at every order in $\tau$ and in $\ndot t$. It needs two numbers that the experiment has, i.e.,
the fraction of runs it kept and the fraction that the closed form says it would have kept under continuous
monitoring. For the flux projector $\Phisec$ one has $R=1$, and for a gauge correlator such as
$\hat\sigma^{x}_{\ell_1}\hat\sigma^{x}_{\ell_2}$ one has $R=2$. For the generators, the bias is zero by
construction. The certificate says that the excess of the measured yield over the closed form is itself the
error budget. For an experimenter, this means that the yield is a diagnostic as well as a cost. For a
population the range is unbounded, because a kept run may contain any even number of gain events on a mode.
Equation~\eqref{eq:cert} with $R$ taken on the sectors reachable with two events, which gives $R=N_0+2$, is
then a leading-order estimate and not a certificate. It fails as a bound when $\ndot t$ is of order one, and
we use it only in the regime of Table~\ref{tab:bias}, where the weight of sectors beyond $N_0+2$ is of order
$(\ndot\tau)^2$.

When the zero-jump state itself differs from the reference, the hypothesis of the theorem fails and the
certificate takes a longer form. This happens for unequal rates and for a mixture of phonon-number sectors
with a common syndrome. Write $Z(t)=\tr\,\Ucal_{t,0}(\Pnom\rho_0\Pnom)$ for the weight of the zero-jump
trajectories that started in the nominal syndrome, computed from the no-jump evolution of
Eq.~\eqref{eq:nojump} at the actual rates, $\rho_{\rm nj}$ for their normalised state and $\alpha=Z/Y$. Then
$\rho_{\rm kept}=\alpha\rho_{\rm nj}+(1-\alpha)\rho_{\rm other}$, where the second term collects the
cancelled pairs, and, for a thermal preparation, the runs that entered the nominal syndrome from outside it.
Hence
\begin{equation}
\big|\tr\hat O\rho_{\rm kept}-\tr\hat O\rho_{\rm ref}\big|\le
\alpha\,\big|\tr\hat O\rho_{\rm nj}-\tr\hat O\rho_{\rm ref}\big|+(1-\alpha)R .
\label{eq:certgen}
\end{equation}
Equation~\eqref{eq:cert} is the case in which the first term vanishes. Both forms are pointwise in time. A
bound on the largest departure within a window is therefore the largest value of the right-hand side within
the window, which is what Table~\ref{tab:bias} reports.

The excess yield can be predicted before the run. To first order in $\tau$ the kept ensemble is the zero-jump
state plus the trajectories with exactly two heating events on the same mode inside one interval. When the
interval is short compared with the hopping time and with the mean time between events, the two events act at
a single instant $s$. The four orderings of gain and loss on a mode give the operators $\hat n_m+1$, $\hat
n_m$, $\hat a_m^{\dagger2}$ and $\hat a_m^{2}$, the double integral over an interval gives $\tau^2/2$, and
the position of the interval is uniform in $[0,t]$. This leads to the following statement, proved in
Appendix~\ref{app:functional}.

\begin{proposition}[First-order bias law and excess yield]
\label{prop:functional}
Let the heating rates be equal, and the initial state have a definite phonon number and a definite syndrome,
and define the two-event map $\Kcal(\rho)=\sum_m\big[(\hat n_m+1)\rho(\hat n_m+1)+\hat n_m\rho\hat
n_m+\hat a_m^{\dagger2}\rho\hat a_m^{2}+\hat a_m^{2}\rho\hat a_m^{\dagger2}\big]$. The bias of the
post-selected estimator of $\hat O$ with syndrome readings every $\tau$ is
\begin{equation}
\tr\hat O\rho_{\rm kept}-\tr\hat O\rho_{\rm ref}=\frac{\ndot^2\tau}{2\,Y_0(t)}\int_0^t\dd s\,
\Big[\tr\hat O\,\Ucal_{t,s}\Kcal\big(\Ucal_{s,0}\rho_0\big)-\tr\hat O\rho_{\rm ref}(t)\;
\tr\Ucal_{t,s}\Kcal\big(\Ucal_{s,0}\rho_0\big)\Big]+O(\tau^2),
\label{eq:functional}
\end{equation}
and the excess yield is $Y-Y_0=\tfrac12\ndot^2\tau\int_0^t\dd s\,\tr\Ucal_{t,s}\Kcal(\Ucal_{s,0}\rho_0)+O(\tau^2)$,
where $\Ucal$ is the no-jump propagator of Eq.~\eqref{eq:nojump}. By Theorem~\ref{thm:exact} every factor is a
heating-free evolution multiplied by a known exponential, so both quantities are computable from the
heating-free dynamics alone. For one phonon on the two modes of the loop, the excess yield is
\begin{equation}
Y-Y_0=Y_0\Big[3\ndot^2\tau t+\ndot\tau\big(1-\ee^{-4\ndot t}\big)\Big]+O(\tau^2).
\label{eq:excess}
\end{equation}
\end{proposition}

The two terms of Eq.~\eqref{eq:excess} have a simple origin. The first comes from the gain-then-loss and
loss-then-gain branches. These return the state to its sector and are damped afterwards at the same rate as
the zero-jump state. The second comes from the double-gain branch, which raises the sector by two and is
damped afterwards at twice the rate. The trace of $\Kcal$ on any one-phonon state of two modes is $14$, split
as $6$ for the returning branches and $8$ for the double gain, and the factor $(1-\ee^{-4\ndot t})/4\ndot$ is
the time integral of the extra damping. The double-gain branch is also why the no-jump propagator in
Eq.~\eqref{eq:functional} cannot be replaced by the heating-free propagator. A first attempt that made that
replacement gave a bias coefficient of $3.7$ instead of $2.0$. At $\ndot t\approx0.35$ the sector-dependent
damping is a large effect and not a refinement.

\begin{figure}
\centering
\includegraphics[width=0.98\textwidth]{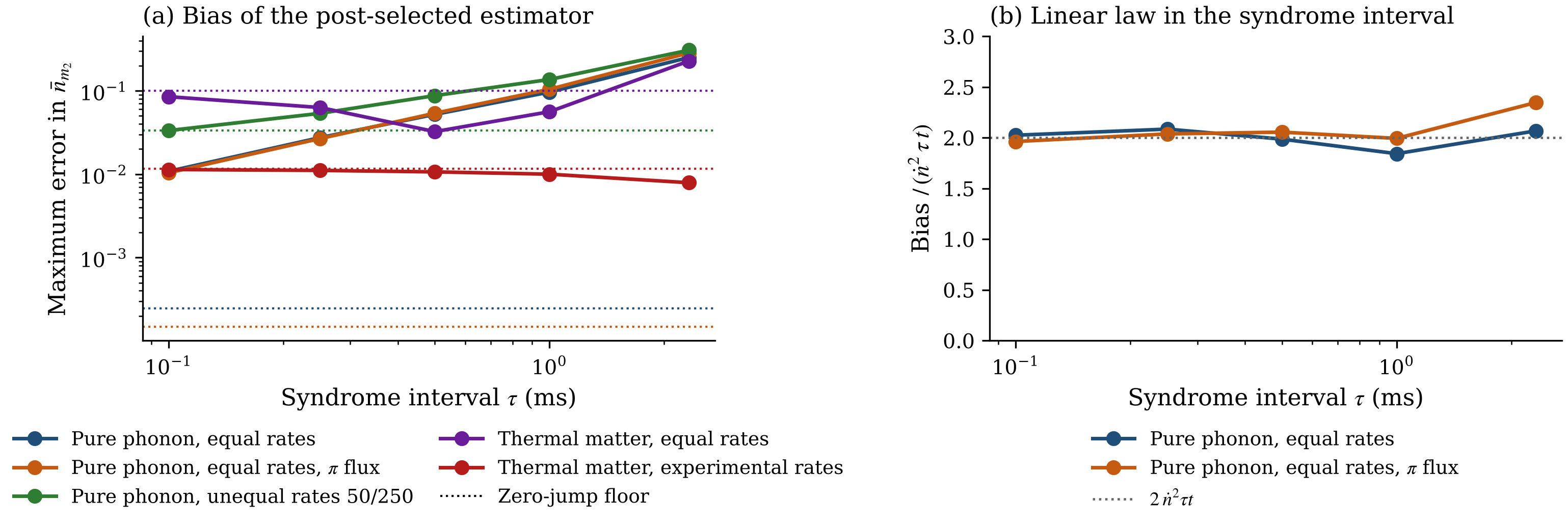}
\caption{Bias of the post-selected estimator against the syndrome interval. (a) The largest departure of the
post-selected population of the second site from the heating-free reference within the window, for five
preparations of the loop. The dotted lines are the errors of the zero-jump conditional state, which the
estimator approaches as the interval shrinks. For one phonon with equal rates, that floor is below
$3\times10^{-4}$. For unequal rates it is the tilt of Section~\ref{sec:unequal}. For thermal matter it is the
floor of Proposition~\ref{prop:thermal}, computed here with the flag construction of
Appendix~\ref{app:numerics} on the ideal phonon-added state, which does not fix the parities. (b) The bias of
the two equal-rate one-phonon cases divided by $\ndot^2\tau t$, with the floor subtracted. The ratio lies
between $1.85$ and $2.10$ for every interval up to $1$ millisecond in both flux sectors and reaches $2.35$ for
the single final check in the $\pi$-flux sector.}
\label{fig:bias}
\end{figure}

\begin{table}
\centering
\caption{The finite-interval bias on the zero-flux loop with one phonon and equal heating rates of $150$
quanta per second at $h/J=0.6$, with $t$ the window of $2.32$ milliseconds. The excess yield is compared with
Eq.~\eqref{eq:excess}. The bias in the flux projector $\Phisec$, which is bounded with $R=1$, is compared
with the certificate of Eq.~\eqref{eq:cert}, reported as its largest value within the window, with the closed
form $Y_0(u)=\ee^{-4\ndot u}$ at every time $u$. The entries are computed at integrator tolerances of
$10^{-10}$ absolute and $10^{-8}$ relative. The bias in the population of the second site is compared with the
law $2\ndot^2\tau t$ obtained from Eq.~\eqref{eq:functional}. All biases are the largest departure within the
window.}
\label{tab:bias}
\small\setlength{\tabcolsep}{4pt}
\begin{tabular}{lccccccc}
\toprule
Interval $\tau$ (ms) & Yield $Y$ & Excess $Y-Y_0$ & Eq.~\eqref{eq:excess} & Bias in $\avg{\Phisec}$ & Certificate & Bias in $\nbar{2}$ & $2\ndot^2\tau t$ \\
\midrule
$2.32$ (final check) & $0.398$ & $0.149$ & $0.155$ & $0.012$ & $0.375$ & $0.251$ & $0.242$ \\
$1.0$ & $0.305$ & $0.057$ & $0.067$ & $0.0083$ & $0.186$ & $0.097$ & $0.104$ \\
$0.5$ & $0.281$ & $0.032$ & $0.034$ & $0.0030$ & $0.114$ & $0.052$ & $0.052$ \\
$0.25$ & $0.265$ & $0.0168$ & $0.0167$ & $0.0014$ & $0.064$ & $0.027$ & $0.026$ \\
$0.1$ & $0.255$ & $0.0068$ & $0.0067$ & $0.00054$ & $0.027$ & $0.011$ & $0.010$ \\
\bottomrule
\end{tabular}
\end{table}

Figure~\ref{fig:bias} and Table~\ref{tab:bias} show the exact bias of the post-selected estimator as the
syndrome interval is varied from the full window, which is a single final check, down to $0.1$ milliseconds.
The excess yield follows Eq.~\eqref{eq:excess} to one percent at the two smallest intervals. At the two
largest it agrees to four and eighteen percent, where the second-order terms in $\tau$ enter. The bias is
linear in $\tau$. For the population of the second site with one phonon and equal rates, the largest bias
within the window with the zero-jump floor subtracted is $c\,\ndot^2\tau t$. The coefficient $c$ lies between
$1.85$ and $2.10$ for every interval up to $1$ millisecond in both flux sectors and for the final check in the
zero-flux sector. For the final check in the $\pi$-flux sector $c=2.35$, which is the elevated point in
Figure~\ref{fig:bias}(b). The round value $2.0$ is therefore accurate to about ten per cent for
$\ndot\tau\le0.15$, and it understates the bias by fifteen per cent at the single final check in the
$\pi$-flux sector, where the second-order terms are largest. The coefficient is a property of the loop at
these parameters and not a universal number, since Eq.~\eqref{eq:functional} depends on the dynamics through
$\rho_s$. Its range of validity in $\ndot\tau$, $J\tau$ and $h/J$ is a matter for a calibrated gate of the
kind used in~\cite{ugail2026noise}, which we have not yet run. The certificate of Theorem~\ref{thm:cert}
holds for the flux projector at every interval. Its margin is large because the cancelled pairs move the flux
only indirectly. In the $\pi$-flux sector, where the frozen state is more sensitive, the bias in
$\avg{\Phisec}$ is $0.070$, $0.030$, $0.014$, $0.0068$ and $0.0027$ at the same five intervals, against
certificates of $0.379$, $0.193$, $0.117$, $0.064$ and $0.027$.

\begin{figure}
\centering
\includegraphics[width=0.6\textwidth]{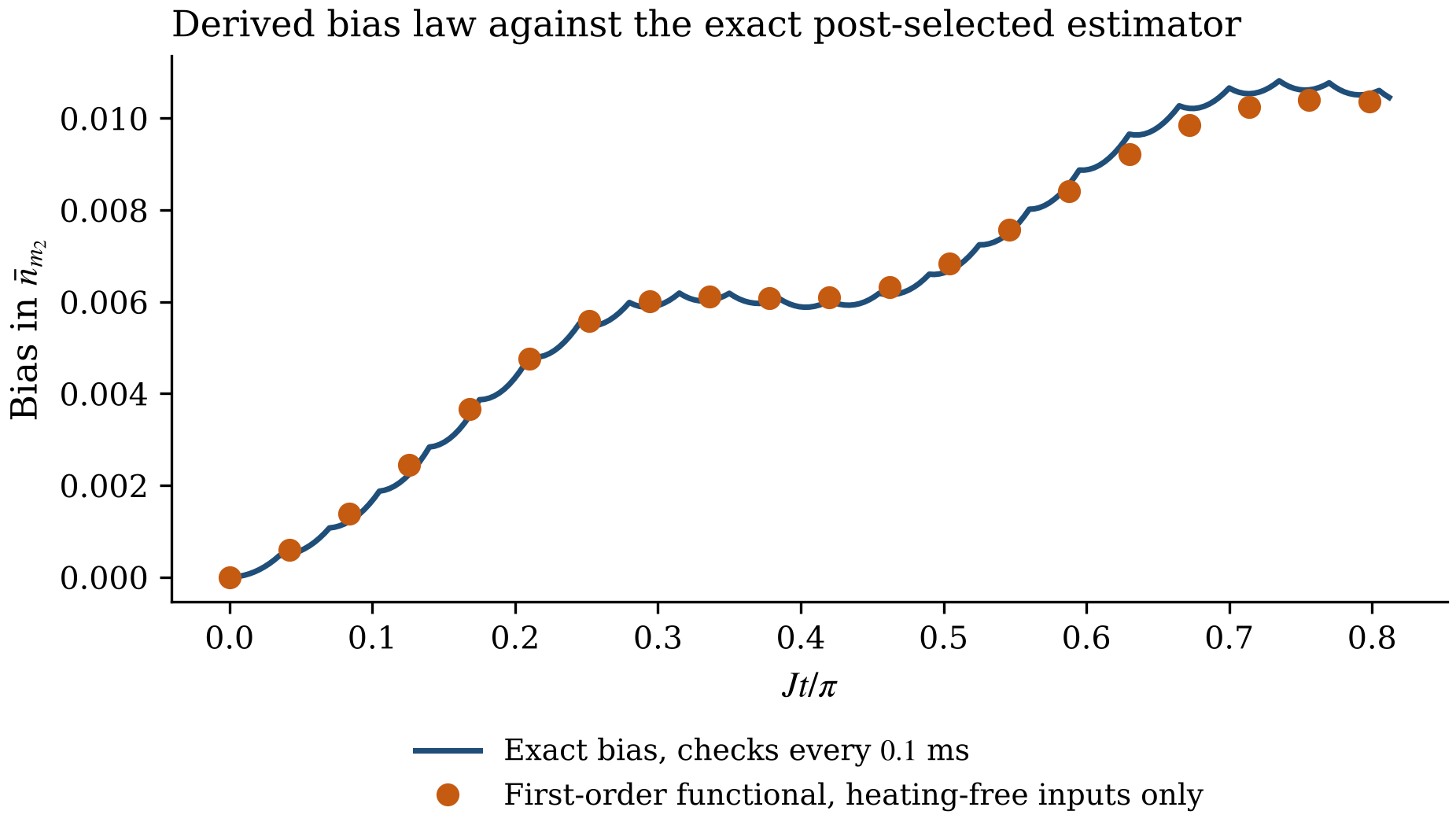}
\caption{The first-order bias law of Proposition~\ref{prop:functional} against the exact bias of the
post-selected estimator with syndrome checks every $0.1$ milliseconds, on the zero-flux loop with one phonon,
equal heating rates of $150$ quanta per second and $h/J=0.6$. The functional uses only heating-free evolutions
and the exponential factors of Theorem~\ref{thm:exact}. It reproduces the full time profile of the exact bias
to $4\times10^{-4}$ and gives a largest bias of $1.99\,\ndot^2\tau t$ against the exact $2.0$.}
\label{fig:functional}
\end{figure}

Figure~\ref{fig:functional} compares the functional with the exact bias at the smallest interval. The
functional reproduces the full time profile to $4\times10^{-4}$ and gives a coefficient of $1.99$ against the
exact $2.0$. The residual of a few percent is the correction of order $J\tau$ from the finite separation of
the two events. The law is of the kind we have elsewhere called a first-order law, derived from the noiseless
dynamics and accurate within a declared margin. It sits beside the certificate of Theorem~\ref{thm:cert},
which is guaranteed and loose, as the certification programme
intends~\cite{ugail2026weyl,ugail2026noise}.

\section{Unequal rates and thermal preparation}
\label{sec:unequal}

The experiment fails two hypotheses of Theorem~\ref{thm:exact}, a uniform bath and a definite phonon number.
We treat them in turn.

With unequal rates, the operator $\hat K$ of Eq.~\eqref{eq:nojump} on the loop is
$(\ndot_1+\ndot_2)(\hat N+1)+\Delta(\hat n_1-\hat n_2)$ with $\Delta=\ndot_1-\ndot_2$. On the sector of one
phonon, its non-constant part is $2\Delta\hat n_1$. This term does not commute with the hopping. The conditional
state is tilted away from the heating-free state, because a trajectory that spent its time on the mode that
heats more slowly is less likely to have been rejected. The hopping limits the tilt, since the phonon does not
stay on either mode for long. At zero electric field and without motional dephasing the tilt has a closed
form.

\begin{proposition}[Tilt of the conditional state]
\label{prop:tilt}
For one phonon on the loop with the link qubits prepared in $\Phip$, at $h=0$, without motional dephasing and
without qubit noise, with the phonon initially on the first site, the population of the second site in the
zero-jump conditional state is
\begin{equation}
\nbar{2}(t)=\frac{(4J^2/\omega^2)\sin^2\omega t}{\big(\cos\omega t-(\Delta/2\omega)\sin\omega t\big)^2+(4J^2/\omega^2)\sin^2\omega t},
\qquad\omega=\sqrt{4J^2-\Delta^2/4},
\label{eq:tilt}
\end{equation}
against the heating-free value $\sin^22Jt$. To first order in $\Delta/J$ the tilt is
$(\Delta/2J)\sin^3(2Jt)\cos(2Jt)$, whose magnitude never exceeds $0.65\,|\Delta|/4J$.
\end{proposition}

The derivation is a two-level calculation. With the link qubits in $\Phip$, the hopping acts with amplitude
$\pm2J$, and the one-phonon subspace is closed under the no-jump evolution. After the removal of the common factor
$\ee^{-(\ndot_1+\ndot_2)t}$, that evolution is generated by a non-Hermitian matrix with off-diagonal entries
$2J$ and diagonal entries $-\ii\Delta/2$ and $+\ii\Delta/2$. Its eigenfrequencies are $\pm\omega$. The
amplitudes from the initial state are $\cos\omega t-(\Delta/2\omega)\sin\omega t$ on the first site and
$-\ii(2J/\omega)\sin\omega t$ on the second, and normalising the squared moduli gives Eq.~\eqref{eq:tilt}.
The tilt vanishes whenever the phonon is entirely on one site, and changes sign every quarter period. It is
largest when the phonon is partly on both sites, because that is when the two components of the state have had
different exposures to the two rates. The frequency $\omega$ differs from $2J$ at second order in $\Delta/J$,
so the first-order form is accurate within the window but not uniformly in time.

\begin{figure}
\centering
\includegraphics[width=0.98\textwidth]{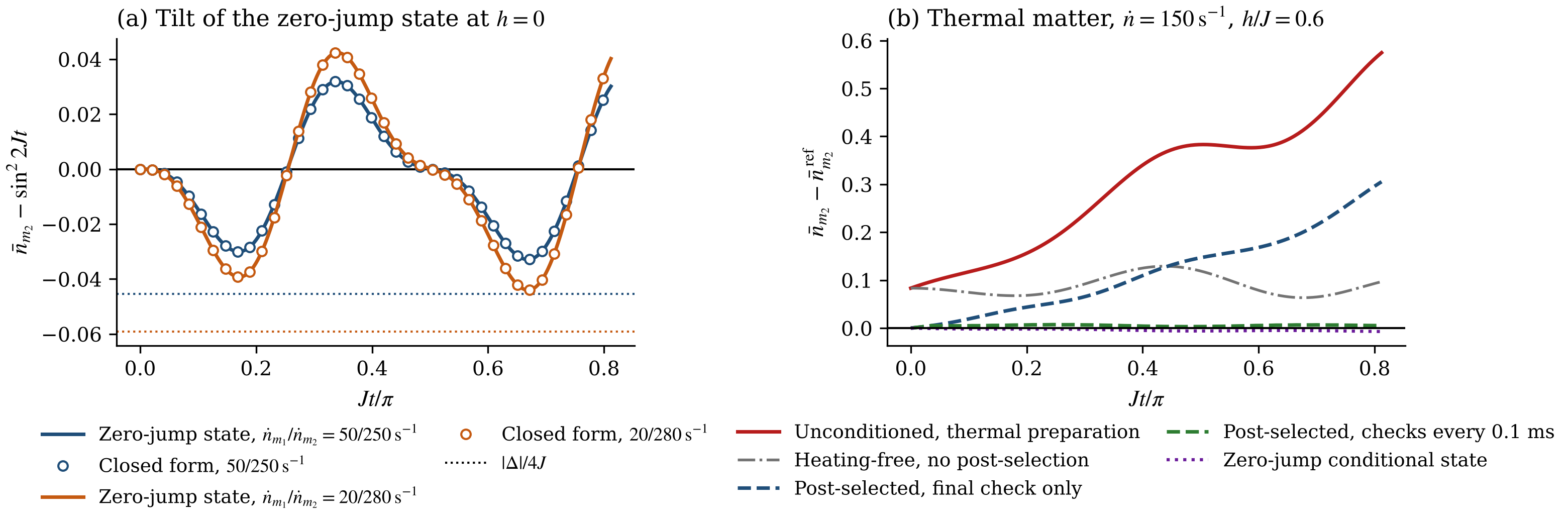}
\caption{The two departures from the hypotheses of Theorem~\ref{thm:exact}. (a) Tilt of the zero-jump
conditional state at $h=0$ without motional dephasing for one phonon on the loop with heating rates of $50$ and
$250$ quanta per second and of $20$ and $280$ quanta per second. The curves are the exact conditional state,
the circles are the closed form of Eq.~\eqref{eq:tilt}, and the dotted lines mark $|\Delta|/4J$. (b) Thermal
matter on the loop with equal heating rates of $150$ quanta per second and $h/J=0.6$, with one phonon added by
a sideband pulse to a thermal background of $\bar n_0=0.1$ on both modes. The reference is the heating-free
dynamics of the component of the prepared state that has the nominal syndrome. The unconditioned evolution
departs from it by up to $0.57$, of which $0.13$ is present without any heating because the thermal background
violates Gauss's law at preparation. Projecting on the nominal syndrome at the final time reduces the
departure to $0.31$, and projecting every $0.1$ milliseconds reduces it to $0.007$. The zero-jump state has a
floor of $0.007$, which is the thermal floor of Proposition~\ref{prop:thermal}.}
\label{fig:tilt}
\end{figure}

Figure~\ref{fig:tilt}(a) compares the closed form with the exact conditional state for rate splits of $50$
and $250$ and of $20$ and $280$ quanta per second. The agreement is $2.4\times10^{-4}$ in both cases. The
largest tilt within the window is $0.72$ and $0.74$ of $|\Delta|/4J$, and the excess over the first-order
value of $0.65$ comes from the second-order terms. With the electric field at $h/J=0.6$ and the experimental
motional dephasing, the population tilt for the split of $50$ and $250$ quanta per second is $0.033$. That is
the floor of the unequal-rate curve in Figure~\ref{fig:bias}(a). The tilt of the flux projector in the same
case is $0.051$, so the flux sector is more sensitive to unequal rates than the population. On the chain with
rates of $50$, $150$ and $250$ quanta per second on the three modes, the population tilt is $0.043$. In every
case computed within the window, the population tilt stayed below $|\Delta|/4J$ evaluated with the largest
rate difference. We report this as an observation about the population, with the first-order form of
Proposition~\ref{prop:tilt} as its analytic basis, and not as a theorem. The flux-projector tilt of $0.051$
exceeds $|\Delta|/4J=0.045$, so the observation does not extend to the flux sector. At the experimental rates
of $5$ and $17$ quanta per second the quantity $|\Delta|/4J$ is $0.003$.

The experiment prepares its matter by a blue-sideband pulse on a thermal background with $\bar n_0=0.1$ on
every mode. We model the pulse as a resonant $\pi$ pulse on the first sideband, calibrated from the ground
state and heralded on the excitation of the ion. It maps the Fock state $|n\rangle$ to
$\sin(\tfrac{\pi}{2}\sqrt{n+1})\,|n+1\rangle$ up to a phase. The prepared state is then a mixture over
phonon-number sectors that is diagonal in the Fock basis. Two things follow from the thermal background. The
state is not an eigenstate of the generators, because the parity of each mode is uncertain, and it has no
definite total phonon number. The syndrome measurement addresses both at once. The nominal syndrome is odd
parity on the site carrying the phonon and even parity on the empty site. A projection onto it retains $0.867$
of the prepared state and fixes the parity of every mode. The total phonon number of the retained part is
therefore odd, with sector weights of $0.990$ on one phonon, $0.0096$ on three and $10^{-4}$ on five. The
thermal background that the projection removes is a violation of Gauss's law present at preparation, not
caused by heating. Relative to the gauge-invariant target it displaces the unconditioned population of the
second site by $0.083$ at $t=0$ and by up to $0.13$ within the window at $h/J=0.6$. The reference for thermal
matter is therefore the heating-free dynamics of the nominal-syndrome component, which is the gauge-invariant
dynamics the experiment is trying to realise. If the pulse is modelled instead as the ideal addition of one
phonon, which weights $|n\rangle\to|n+1\rangle$ by $\sqrt{n+1}$, the retained fraction is $0.777$, and the
sector weights are $0.967$, $0.032$ and $6\times10^{-4}$. The conclusions below are insensitive to the
preparation model, although the yield is not.

\begin{proposition}[Thermal preparation]
\label{prop:thermal}
For an initial state $\rho_0=\sum_Np_N\rho_N$ that is block diagonal in the phonon-number sectors and has a
definite syndrome, the zero-jump conditional state under equal rates is $\sum_Nw_N(t)\,\ee^{t\Lcal_0}\rho_N$
with $w_N(t)\propto p_N\ee^{-2\ndot Nt}$. For a bounded observable of range $R$, the departure from the
heating-free value is bounded by $R\cdot\mathrm{TV}(w,p)$, where $\mathrm{TV}$ is the total-variation distance
between the two sets of weights, which is computable from the preparation model and $\ndot t$ alone. If the
total phonon number of each kept run were known, weighting the run by $\ee^{2\ndot Nt}$ would restore the
weights $p_N$ and the heating-free estimator exactly on the zero-jump ensemble.
\end{proposition}

The first part is Theorem~\ref{thm:exact} applied to each block. The bound follows from writing the departure
as $\sum_N(w_N-p_N)\tr\hat O\ee^{t\Lcal_0}\rho_N$ and subtracting the midpoint of the range. For a population
the range is unbounded, and the same decomposition gives instead the number-weighted bound
$\max\{\sum_NN(w_N-p_N)_+,\sum_NN(p_N-w_N)_+\}$, since $0\le\avg{\hat n_m}_N\le N$ in the sector $N$. With
the sector weights fixed by the nominal projection, the total-variation distance at the end of the window is
$0.0072$ at $150$ quanta per second, and the number-weighted bound is $0.022$. The thermal floor is therefore
small once the syndrome has been projected. Figure~\ref{fig:tilt}(b) shows the loop with equal rates of $150$
quanta per second at $h/J=0.6$. The zero-jump state departs from the reference by $0.0072$ on the population,
which is a computed value and not a certified one, and by $0.0007$ on the flux projector, within its
certificate. At the experimental rates, the same quantities are a total-variation distance of $0.0009$ and a
number-weighted bound of $0.0028$, evaluated at the homogeneous comparison rate of $11$ quanta per second,
which has the same total rate as the actual $5$ and $17$. The reweighting in the last part of the proposition
is exact on the zero-jump ensemble. We do not recommend it for a protocol with a finite interval, because the
weight $\ee^{2\ndot Nt}$ amplifies the double-gain trajectories that the finite interval admits. At the
experimental rates with a single final check it changes the population error from $0.011$ to $0.013$, and at
$150$ quanta per second it makes the estimator worse by an order of magnitude. Since the nominal projection
already leaves a computed floor below $10^{-3}$ at the experimental rates, the reweighting is not needed. A
Fock-state preparation by a resolved sideband sequence would remove the floor entirely by making $p_N$ a
delta.

\section{Two readout protocols}
\label{sec:readout}

The results of Sections~\ref{sec:exact} to~\ref{sec:unequal} assume that the syndrome can be read. A
destructive measurement at the end of a run, given a readout that resolves the phonon number of one mode, can
test some of them. A non-demolition measurement during the run is needed for the rest. We state the two
separately, with the results that each supports.

The first protocol is a destructive measurement of the Gauss generators at the final time only, followed by
acceptance of the runs whose outcome is the nominal syndrome. It needs no record of the initial syndrome,
because the gauge-invariant dynamics preserves the syndrome. A final outcome equal to the nominal one is
exactly the acceptance rule of Section~\ref{sec:finite} with $\tau=t$ when the initial syndrome is definite,
and the projection of Section~\ref{sec:unequal} when it is not. The ingredients are the primitives the device
already uses, together with one it has not demonstrated, a number-resolving readout of a mode. Their sequence
is more demanding than any single measurement the experiment reports. We describe it in order to be clear
about what is assumed and to show that the accepted ensemble is the one simulated.

\begin{figure}
\centering
\includegraphics[width=0.98\textwidth]{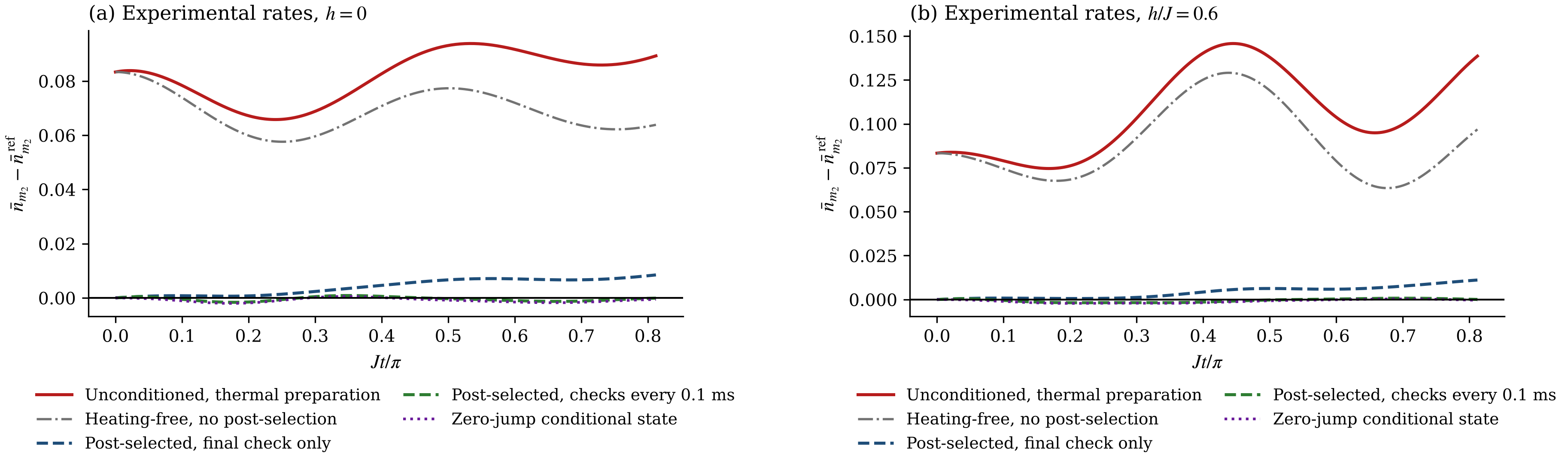}
\caption{Sector tracking at the reported noise of the experiment, with heating rates of $5$ and $17$ quanta per
second, the experimental coherence times and matter prepared by a sideband pulse on a thermal background with
$\bar n_0=0.1$. The curves show the departure of the population of the second site from the gauge-invariant
reference, which is the heating-free dynamics of the nominal-syndrome component of the prepared state, for (a)
$h=0$ and (b) $h/J=0.6$. The unconditioned evolution is displaced by the thermal background that violates
Gauss's law at preparation, shown by the heating-free curve without post-selection, and further by the
heating. A single Gauss-law measurement at the final time with acceptance of the nominal syndrome reduces the
largest departure from $0.094$ to $0.0085$ at $h=0$ and from $0.146$ to $0.011$ at $h/J=0.6$, keeping $0.80$ of
the runs. Syndrome readings every $0.1$ milliseconds reduce it to $0.0017$ and $0.0018$, close to the
unequal-rate floor of the zero-jump state.}
\label{fig:exp}
\end{figure}

The link qubits are read in the $\hat\sigma^{x}$ basis by a $\pi/2$ rotation followed by fluorescence
detection. Fluorescence preserves the motional state of an ion only for the dark outcome~\cite{saner}, so any
readout of the modes that follows is available only on the branch in which both ions are dark. That branch
cannot be treated as a simple readout loss, because which $\hat\sigma^{x}$ outcome pair is dark is correlated
with the matter. At $h=0$ the zero-flux loop evolves as
$\cos(2Jt)\,|1,0\rangle|\Phi^{+}\rangle-\ii\sin(2Jt)\,|0,1\rangle|\Phi^{-}\rangle$. The state
$|\Phi^{+}\rangle$ has equal $\hat\sigma^{x}$ outcomes on the two links and $|\Phi^{-}\rangle$ has opposite
ones. Keeping the branch in which both ions are dark under a single readout setting would therefore keep only
the component with the phonon on the first site and estimate a population of zero. The remedy is to use four
readout settings, which differ by a $\pi$ phase on the $\pi/2$ rotation of one or both ions, so that each of
the four outcome pairs $(\pm,\pm)$ is the dark pair in one setting. In each setting, the dark branch is followed
by a reset of the qubits and a readout of the modes. The accepted counts and the accepted sums of the matter
observable are added over the four settings before the ratio is formed. The four settings are given the same
number of attempted shots, or else each setting's counts and sums are divided by its number of attempted shots
before the addition. The four outcome projectors $\hat P_{\pm\pm}$ sum to the identity, commute with $\Pnom$,
and commute with every observable of the modes. The combined estimator is therefore
$\sum_{\pm\pm}\tr(\hat P_{\pm\pm}\Pnom\hat O\rho)/\sum_{\pm\pm}\tr(\hat P_{\pm\pm}\Pnom\rho)=\tr(\Pnom\hat
O\rho)/\tr(\Pnom\rho)$, which is the estimator of Figure~\ref{fig:exp}, obtained at the cost of four times the
shots of an ideal non-destructive syndrome measurement.

The nominal syndrome in each setting is the pair of parities that gives $\Gm{1}=-1$ and $\Gm{2}=+1$ for the
known value of $\hat\sigma^{x}_{\ell_1}\hat\sigma^{x}_{\ell_2}$ on that branch. The readout of the modes must
supply the parity of the first mode, the parity of the second mode and the population of the second mode in
the same shot. A readout that resolves the phonon number of the second mode provides this, together with a
parity mapping of the first mode onto its qubit~\cite{leibfried,fluhmann}. A consistent ordering on the dark
branch is as follows. Record the known product $s=\hat\sigma^{x}_{\ell_1}\hat\sigma^{x}_{\ell_2}$ of that
branch. Map the required parity $\hat P_1=-s$ of the first mode onto its qubit and reject the complement. Then
measure the phonon number of the second mode and keep the outcomes with $(-1)^{n_2}=s$. The last and
destructive step then comes after every quantity the estimator needs has been recorded. Projective
phonon-number measurement has been demonstrated in trapped ions by repeated sideband
interrogation~\cite{an2015}, although not on this device. With only a parity bit per mode per shot the
population could not be read in the same shot as the syndrome. For the flux projector, the individual
$\hat\sigma^{x}$ readouts do not commute with $\hat W_C=\hat\sigma^{z}_{\ell_1}\hat\sigma^{z}_{\ell_2}$. The
flux does commute with the product $\hat\sigma^{x}_{\ell_1}\hat\sigma^{x}_{\ell_2}$ that the syndrome needs.
A readout of the link qubits in the Bell basis, by the entangling gate the experiment uses for preparation
followed by fluorescence, therefore gives both jointly, with the same four-setting reconstruction of the dark
branch. The yields quoted below are the acceptance probabilities of the syndrome rule, which the four-setting
reconstruction recovers in full at the cost of the extra shots.

Figure~\ref{fig:exp} shows what the first protocol would do at the reported noise of the experiment, with
heating rates of $5$ and $17$ quanta per second, the experimental coherence times and the thermal preparation
of Section~\ref{sec:unequal}. At $h/J=0.6$ the population of the second site departs from the gauge-invariant
reference by up to $0.146$ within the window. Of this, $0.129$ is the thermal background present at
preparation, and the rest is heating, which on its own displaces the population by up to $0.042$. A single
final measurement with acceptance of the nominal syndrome reduces the departure to $0.011$ while keeping
$0.80$ of the runs. The flux projector is recovered to $0.002$, against a certificate from
Eq.~\eqref{eq:certgen} of $0.019$ with $\alpha=0.984$ at the actual rates. At $h=0$ the numbers are $0.094$,
$0.0085$ and $0.79$. The bias law of Section~\ref{sec:finite} gives $2\ndot^2\tau t\approx10^{-3}$ for the
final check at these rates. The rest of the residual is contamination from runs that started outside the
nominal syndrome and reached it by an odd number of events, which a final-only measurement cannot distinguish
from runs that never left it. With syndrome readings every $0.1$ milliseconds the departure falls to $0.0018$
at $h/J=0.6$ and to $0.0017$ at $h=0$. The zero-jump floor is $0.002$, which is the unequal-rate tilt of the
modes at $5$ and $17$ quanta per second and is consistent with $|\Delta|/4J=0.003$. The thermal floor of
Proposition~\ref{prop:thermal} is $0.0009$ at these rates and is not the limiting term.

The second protocol is an ancilla-assisted non-demolition measurement of $\hat G_m$ during the run. It
requires a probe ion coupled to the mode and to the link qubits, so that the parity of the mode and the
$\hat\sigma^{x}$ values of the links can be mapped onto the probe and read without disturbing the gauge
field. The link qubits themselves are the gauge field and cannot be read mid-run without collapsing it.
Mid-circuit measurement of an ancilla, with the data ions shielded from the readout light, is established in
the quantum-charge-coupled-device architecture~\cite{pino}. Parity mapping of a motional mode onto an ion is
established in bosonic encodings~\cite{fluhmann}. The ingredients therefore exist, although their combination
on this device does not. This protocol supports the full interval scan of Table~\ref{tab:bias}, the bias law
of Proposition~\ref{prop:functional}, and the full strength of Theorem~\ref{thm:exact}. The interval scan is
the quantitative case for building it. At the heating rates of the present device, the first protocol captures
most of the benefit. The second becomes necessary when the heating rates or the evolution times grow, as they
will on the larger geometries of the roadmap of~\cite{saner}, because the bias of a single final check grows
as $\ndot^2t^2$ while the bias with checks every $\tau$ grows as $\ndot^2\tau t$.

\section{Discussion}
\label{sec:discussion}

The main practical statement of the paper concerns a single measurement of the Gauss generators at the end of
each run, with acceptance of the nominal syndrome. It removes most of the heating damage predicted
in~\cite{ugail2026noise}. It also removes the part of the thermal preparation that violates Gauss's law.
At the reported rates, the departure of the population from the gauge-invariant target falls from $0.146$ to
$0.011$ at a yield of $0.80$, and the residual is understood. It is the sum of four terms. There is a
finite-interval bias of order $10^{-3}$ with a derived coefficient, an unequal-rate tilt of order
$2\times10^{-3}$ with a closed form at zero field, and a computed thermal floor below $10^{-3}$ whose
certified bound at the homogeneous comparison rate is $3\times10^{-3}$. The fourth term is contamination from
runs that entered the nominal syndrome by an odd number of events, which a final-only measurement cannot
exclude and which syndrome readings during the run would remove. Three consequences are useful to an
experimenter. The final Gauss-law measurement should become a standard part of the protocol. The fraction of
runs it rejects should be compared with the zero-jump weight $Z(t)$ of the nominal component, computed at the
actual rates from the heating-free dynamics, as a check on the heating and preparation models. And the
certificate of Eq.~\eqref{eq:certgen} turns the measured yield into an error bound for every bounded
observable, once the small zero-jump bias at the actual rates has been computed.

The theoretical statement is narrower and should be placed with care. Post-selection on a symmetry is
approximate in general, because the no-jump evolution of a non-unitary channel distorts the surviving state.
The distortion is absent whenever the total jump rate depends on a conserved quantity alone. That mechanism is
known. It is why fixed-excitation bosonic codes survive amplitude damping~\cite{chuang}, and it is the
defining property of a uniformly decaying subspace~\cite{suri}. For matter encoded in spins under Pauli noise,
the question does not arise, because the jump operators are unitary, and the symmetry-verification literature
for such matter is concerned with which errors the syndrome detects rather than with what the surviving state
looks like~\cite{froland,ballini,modi}. The present paper identifies the phonon number of the gauge theory as
the conserved quantity and the Gauss generators as a syndrome that reads the heating history without
disturbing the dynamics. Above all, it analyses what happens when the conditions of the mechanism are relaxed.
The heating channel produces return events that pure loss does not at low occupation, and the bias and excess
yield they cause have been derived in closed form and verified on exact dynamics. The unequal-rate tilt and
the thermal preparation have been treated in the same way. These parts of the paper have no counterpart in
the uniformly decaying subspace construction, which excludes return events by hypothesis, and they would be
needed for pure loss too as soon as a mode holds two quanta.

The limitations are those of the model, the protocol and the numerics. The heating model is an
infinite-temperature bath, which is the standard model for electrode noise and the model the experiment
reports against. Theorem~\ref{thm:exact} does not depend on the temperature of the bath as long as it is the
same on every mode. A mode dependence of the total rate produces a tilt of the form of
Proposition~\ref{prop:tilt}. The motional dephasing is Markovian, as in~\cite{ugail2026noise}, and the method
is indifferent to its form because the syndrome does not see it. The preparation is modelled as a heralded
sideband $\pi$ pulse on a thermal state. The nominal fraction of the preparation, which sets the yield, depends
on this model, while the sector weights after the projection do not depend on it appreciably. The readout
sequence of the first protocol has been shown to reproduce the simulated estimator through the four-setting
reconstruction, but it has not been simulated with readout errors, and it requires a number-resolving readout
of one mode that this device has not demonstrated. The numerics are exact Lindblad evolutions. Their residuals
converge with the integrator tolerance, and the post-selected quantities change in the third decimal between
Fock dimensions six and eight. The finite-interval scan was carried out on the loop only. The theorem and the
functional are free of geometry. The bound $|\Delta|/4J$ on the tilt is an observation within the window with
an analytic basis at first order, not a theorem. The bias law has been verified on exact dynamics but not yet
admitted through a calibrated gate of the kind used in~\cite{ugail2026noise}. Such a gate would fix the range
of $\ndot\tau$, $J\tau$ and $h/J$ over which the coefficient may be trusted, and it is the natural next step
before the law is used to design an interval.

\section{Conclusions}
\label{sec:conclusions}

In this work, we have asked whether the damage that heating does to a hybrid qubit--oscillator quantum
simulation of a $\ZZ$ lattice gauge theory can be removed rather than only predicted. We have found that it
can be removed by a measurement of the Gauss generators at the end of each run. The measurement is implementable with
the primitives of the device, together with a number-resolving readout of one mode and a four-setting
reconstruction of the destructive readout. The generators are phonon parities. Heating flips them, and the
gauge-invariant dynamics does not, so the generators form a syndrome that records the heating history of a run
without disturbing the quantities the experiment wants to measure. Keeping the runs whose syndrome never
changed gives an estimator that equals the heating-free estimator at every order, for a bath whose total rate
is the same on every mode, for a state of definite phonon number and in the limit of continuous monitoring.
Its yield has the closed form $\ee^{-(sN+C)t}$, which for the heating model with equal gain and loss is
$\ee^{-\ndot(2N+M)t}$. The mechanism is the known one by which a fixed excitation-number sector decays
uniformly. What the gauge theory supplies is the conserved number and the syndrome that reads it.

This paper is about what happens when the conditions are relaxed. When the syndrome is read at
finite intervals the heating channel admits runs in which two events cancelled, at any occupation rather than
only at two quanta or more as for pure loss, and the estimator acquires a bias that is linear in the interval.
We have derived its coefficient and the accompanying excess yield from the heating-free dynamics and verified
both against exact Lindblad evolutions. We have given a certificate for the bias of any bounded observable.
For a state of definite phonon number it needs only the measured yield and its closed form, and otherwise it
needs in addition the zero-jump bias at the actual rates. Unequal heating rates tilt the conditional state by
an amount with a closed form at zero field, and the population tilt stayed below the rate difference over the
hopping in every case computed. Thermal preparation is handled by the same measurement. Projecting on the
nominal syndrome fixes the parity of every mode and removes the part of the background that violates Gauss's
law at preparation. The floor it leaves is computed to be below $10^{-3}$ at the reported rates and certified
to be below $3\times10^{-3}$ at the homogeneous comparison rate.

At the reported noise of the experiment, a single Gauss-law measurement at the end of each run reduces the
departure of the matter population from its gauge-invariant target from $0.146$ to $0.011$. The acceptance is
eight runs in ten, which the four-setting readout recovers at a fourfold cost in shots. Syndrome readings
during the run would reduce the departure further, to the unequal-rate floor of $0.002$. The interval scan is
the quantitative case for a non-demolition measurement of the generators on future hardware, and such a
measurement will become necessary as heating rates and evolution times grow on larger geometries. Thus, with~\cite{ugail2026noise}, which predicts what breaks, the work presented here shows how to recover it. The two leave motional
dephasing and qubit bit flips as the noise processes that a gauge-theory experiment on this platform must
still live with.

\section*{Acknowledgments}
The author would like to thank the Centre for Visual Computing and Intelligent Systems at the University of
Bradford for providing the computing resources used in this work.

\section*{Funding}
No specific funding was received for this work.

\appendix

\section{Derivation of the first-order bias law and the excess yield}
\label{app:functional}

Let the window $[0,t]$ be divided into intervals $I_k=[(k-1)\tau,k\tau]$, and let the heating jump operators
on mode $m$ be $\hat L_{m,-}=\sqrt{\ndot}\,\hat a_m$ and $\hat L_{m,+}=\sqrt{\ndot}\,\hat a_m^{\dagger}$. In
the quantum-jump unravelling, the unnormalised state of the trajectories with jumps $(\hat L_1,u_1),\dots,(\hat
L_n,u_n)$ at ordered times is $\Ucal_{t,u_n}\mathcal J_n\Ucal_{u_n,u_{n-1}}\cdots\mathcal J_1\Ucal_{u_1,0}\rho_0$
with $\mathcal J\rho=\hat L\rho\hat L^{\dagger}$. The state of the full ensemble is the sum over histories
with the time integrals~\cite{dalibard,plenio}. The kept ensemble is the sum over histories in which the
number of jumps on every mode inside every interval is even. To first order in $\tau$ it contains the
zero-jump history and the histories with exactly two jumps, both on the same mode and inside the same
interval. For two jumps on mode $m$ at times $u_1<u_2$ in $I_k$ the contribution is
\begin{equation}
\sum_{\alpha,\beta=\pm}\int_{I_k}\dd u_2\int_{(k-1)\tau}^{u_2}\dd u_1\;
\Ucal_{t,u_2}\mathcal J_{m,\beta}\Ucal_{u_2,u_1}\mathcal J_{m,\alpha}\Ucal_{u_1,0}\rho_0 .
\end{equation}
When $\tau$ is short compared with $1/J$, $1/\ndot$ and the dephasing times, the propagator $\Ucal_{u_2,u_1}$
between the two jumps may be replaced by the identity, and both jumps placed at a common time $s$. The sum over
the four orderings gives
$\sum_{\alpha,\beta}\hat L_{m,\beta}\hat L_{m,\alpha}\rho\hat L_{m,\alpha}^{\dagger}\hat
L_{m,\beta}^{\dagger}=\ndot^2\big[(\hat n_m+1)\rho(\hat n_m+1)+\hat n_m\rho\hat n_m+\hat a_m^{\dagger2}\rho\hat
a_m^{2}+\hat a_m^{2}\rho\hat a_m^{\dagger2}\big]$, which is $\ndot^2$ times the $m$ term of $\Kcal(\rho)$. The
double time integral over the interval gives $\tau^2/2$. Summing over the $t/\tau$ intervals turns the sum over
$k$ into an integral $\int_0^t\dd s/\tau$, with the common time $s$ uniform in $[0,t]$. The unnormalised kept
state is therefore
\begin{equation}
\tilde\rho_{\rm kept}=\Ucal_{t,0}\rho_0+\frac{\ndot^2\tau}{2}\int_0^t\dd s\,\Ucal_{t,s}\Kcal\big(\Ucal_{s,0}\rho_0\big)+O(\tau^2).
\end{equation}
Its trace is the yield, which gives the excess of Proposition~\ref{prop:functional}, and the post-selected
estimator is $\tr\hat O\tilde\rho_{\rm kept}/\tr\tilde\rho_{\rm kept}$. Expanding the ratio to first order in
$\tau$, with $\tr\Ucal_{t,0}\rho_0=Y_0(t)$ and $\tr\hat O\Ucal_{t,0}\rho_0=Y_0(t)\tr\hat O\rho_{\rm ref}(t)$ by
Theorem~\ref{thm:exact}, gives Eq.~\eqref{eq:functional}. The corrections of order $\tau^2$ come from the
finite separation of the two jumps, from histories with four jumps, and from the second-order term of the
ratio.

For one phonon on two modes the branch $\hat a_m^{2}\rho\hat a_m^{\dagger2}$ vanishes. The traces of the
remaining branches are $\sum_m\avg{(\hat n_m+1)^2}=5$, $\sum_m\avg{\hat n_m^2}=1$ and $\sum_m\avg{(\hat
n_m+1)(\hat n_m+2)}=8$ for every state in the sector, and they sum to $14$. The first two branches leave the
state in the one-phonon sector, where $\Ucal_{t,s}$ is the heating-free evolution times $\ee^{-4\ndot(t-s)}$.
With $\tr\Ucal_{s,0}\rho_0=\ee^{-4\ndot s}$ they contribute $\tfrac12\ndot^2\tau\cdot6\cdot t\,\ee^{-4\ndot
t}$ to the excess yield. The double-gain branch puts the state in the three-phonon sector, where
$\Ucal_{t,s}$ carries $\ee^{-8\ndot(t-s)}$, and it contributes $\tfrac12\ndot^2\tau\cdot8\int_0^t\ee^{-4\ndot
s}\ee^{-8\ndot(t-s)}\dd s=\ndot\tau\,\ee^{-4\ndot t}(1-\ee^{-4\ndot t})$. Together these give
Eq.~\eqref{eq:excess}. In the evaluation of Eq.~\eqref{eq:functional} on the loop, the integrand was computed
from the unnormalised no-jump states $\Ucal_{s,0}\rho_0$ on a grid of twenty source times. Each
$\Kcal(\Ucal_{s,0}\rho_0)$ was propagated under the no-jump Lindbladian of Eq.~\eqref{eq:nojump} to every
later time, and the integral was taken by the trapezoidal rule.

\section{Numerical constructions, truncations and convergence}
\label{app:numerics}

A post-selected estimator is a ratio of expectation values. A Monte Carlo evaluation by sampled trajectories
would carry a statistical error large enough to obscure the residuals reported here, so two exact
constructions are used instead. For a prepared state of definite syndrome we attach to each mode a flag
qubit, initially in $|0\rangle$, and replace each heating jump operator $\hat L$ on mode $m$ by $\hat L\hat
X_m$, where $\hat X_m$ is the Pauli $X$ operator on the flag of mode $m$. The Lindblad evolution on the
enlarged space is deterministic and exact, and every heating event on mode $m$ flips its flag. The flag
operators commute with everything else. The block of the state with all flags in $|0\rangle$ is therefore
exactly the unnormalised state of the trajectories with an even number of heating events on every mode since
the flags were last reset. A syndrome reading at time $k\tau$ is implemented by projecting the flags onto
$|0\rangle$ with the projector $\hat\Pi_0$, which discards the rejected trajectories, and continuing the
evolution. The post-selected estimator at any time is $\tr(\hat\Pi_0\hat O\rho)/\tr(\hat\Pi_0\rho)$ and the
yield is $\tr(\hat\Pi_0\rho)$. This is the acceptance rule of a syndrome compared with a known initial
syndrome. For the thermal preparation, the initial syndrome is not definite, and the physical rule is acceptance
of the nominal syndrome at every reading. That rule is simulated without flags, by evolving under the full
Lindbladian and applying the projector $\Pnom=\tfrac14(1-\Gm{1})(1+\Gm{2})$ at every reading, with the
estimator $\tr(\Pnom\hat O\rho)/\tr(\Pnom\rho)$ within each interval. The two constructions coincide for a
state of definite syndrome. They differ for the thermal preparation, where the second also removes the
component of the prepared state that violates Gauss's law. The flag projector commutes with every observable
of the matter and gauge sectors. The physical projector $\Pnom$ commutes with every gauge-invariant
observable, which includes every observable considered in this paper, but not with operators such as $\hat
a_m$ that change the syndrome. Commutation means that the reading does not disturb the observable on the kept
runs. It does not mean that the selection leaves the expectation unchanged, as the thermal projection and the
readout branches of Section~\ref{sec:readout} show.

The loop calculations use two link qubits and two modes with a Fock space of dimension six, plus two flags in
the flag construction. The space has dimension $576$ with flags and $144$ without. The evolution is carried
out with the master-equation solver of QuTiP~\cite{qutip} at the default tolerances of $10^{-8}$ absolute and
$10^{-6}$ relative unless stated. The reference, the unconditioned evolution and the zero-jump conditional
state are computed on the same space from the heating-free Lindbladian, the full Lindbladian and the no-jump
Lindbladian of Eq.~\eqref{eq:nojump} respectively. The post-selected estimators are computed for syndrome
intervals of $1$, $0.5$, $0.25$ and $0.1$ milliseconds and for a single final check. The chain and triangle
calculations of Table~\ref{tab:geometries} use the no-jump Lindbladian directly, with a Fock dimension of two
for the conditional state and four for the unconditioned damage, on spaces of dimension $128$ and $1024$ for
the chain and $64$ and $512$ for the triangle. In the dimension-two calculations the rate operator is
constructed from the analytic form $\hat K=\sum_m\ndot_m(2\hat n_m+1)$ and not from truncated ladder
operators. On a two-dimensional Fock space $\hat a\hat a^{\dagger}$ differs from $\hat n+1$ on the upper
state and would give the wrong yield.

Theorem~\ref{thm:exact} provides an exact benchmark for the integrator, and Table~\ref{tab:conv} reports it.
On the loop without flags, the departure of the zero-jump population from the reference is $6\times10^{-5}$ at
the default tolerances. It falls to $6\times10^{-7}$ when the tolerances are tightened by two orders of
magnitude and to $1\times10^{-8}$ when they are tightened by four, with the same values at Fock dimensions six
and eight. The departure of the yield from the closed form falls from $10^{-10}$ to $10^{-15}$. The
residuals quoted in the body for the zero-jump state at the default tolerances are therefore integrator error
and not truncation error. This benchmark does not test the truncation of the heated trajectories, which reach
higher sectors. The post-selected and unconditioned quantities of Table~\ref{tab:bias} were therefore
recomputed at Fock dimension eight at the tighter tolerances, by the nominal-syndrome projection, which
coincides with the flag construction for this state. Table~\ref{tab:trunc} reports the comparison. The
unconditioned damage changes by $0.004$, the final-check yield by $0.0006$, the final-check population bias
by $0.0005$ and the population bias with checks every $0.1$ milliseconds by $5\times10^{-5}$. The
flux-projector biases change in the third significant figure. The entries of Table~\ref{tab:bias} are
therefore correct to the precision quoted, except that the final-check yield rounds to $0.397$ at dimension
eight. A truncated gain operator changes the rate operator $\hat K$ at the upper boundary of the Fock space,
and these differences are its effect at $\ndot t\le0.35$. The thermal preparation is the heralded sideband map
of Section~\ref{sec:unequal} applied to a thermal state with $\bar n_0=0.1$ on the first mode, with a thermal
state on the second. The sector projectors needed for Proposition~\ref{prop:thermal} are diagonal in the Fock
basis and are applied as observables. The final interval of a syndrome scan ends at the final time whether or
not it has the full length $\tau$. The evolution is carried out in units where $J=1$. The complete scripts,
together with the results files from which every number in the paper is read, are available from the author.

\begin{table}
\centering
\caption{Truncation study of the heated and post-selected calculations on the zero-flux loop with one phonon,
equal heating rates of $150$ quanta per second, the experimental coherence times and $h/J=0.6$, at integrator
tolerances of $10^{-10}$ absolute and $10^{-8}$ relative. The post-selected quantities are computed by the
nominal-syndrome projection, which coincides with the flag construction for this state. Biases are the largest
departure within the window.}
\label{tab:trunc}
\small
\begin{tabular}{lcc}
\toprule
Quantity & Fock dimension $6$ & Fock dimension $8$ \\
\midrule
Unconditioned damage to $\nbar{2}$ & $0.4961$ & $0.4997$ \\
Final-check yield & $0.3976$ & $0.3970$ \\
Final-check bias in $\nbar{2}$ & $0.2508$ & $0.2512$ \\
Final-check bias in $\avg{\Phisec}$ & $0.01237$ & $0.01233$ \\
Yield with checks every $0.1$ ms & $0.25535$ & $0.25534$ \\
Bias in $\nbar{2}$ with checks every $0.1$ ms & $0.01070$ & $0.01065$ \\
Bias in $\avg{\Phisec}$ with checks every $0.1$ ms & $0.000536$ & $0.000534$ \\
\bottomrule
\end{tabular}
\end{table}

\begin{table}
\centering
\caption{Convergence of the zero-jump benchmark on the zero-flux loop with one phonon, equal heating rates of
$150$ quanta per second, the experimental coherence times and $h/J=0.6$, without flags. The population error
is the largest departure of the zero-jump conditional population of the second site from the heating-free
reference within the window, and the yield error is the departure of the computed no-jump weight from
$\ee^{-4\ndot t}$ at the end of the window.}
\label{tab:conv}
\small
\begin{tabular}{ccccc}
\toprule
Fock dimension & Absolute tolerance & Relative tolerance & Population error & Yield error \\
\midrule
$6$ & $10^{-8}$ & $10^{-6}$ & $6.5\times10^{-5}$ & $2.4\times10^{-10}$ \\
$6$ & $10^{-10}$ & $10^{-8}$ & $6.1\times10^{-7}$ & $1.4\times10^{-12}$ \\
$6$ & $10^{-12}$ & $10^{-10}$ & $1.1\times10^{-8}$ & $3.4\times10^{-15}$ \\
$8$ & $10^{-8}$ & $10^{-6}$ & $4.6\times10^{-5}$ & $2.8\times10^{-10}$ \\
$8$ & $10^{-10}$ & $10^{-8}$ & $5.3\times10^{-7}$ & $2.2\times10^{-12}$ \\
$8$ & $10^{-12}$ & $10^{-10}$ & $1.0\times10^{-8}$ & $5.0\times10^{-15}$ \\
\bottomrule
\end{tabular}
\end{table}

\end{document}